\documentclass[article,twocolumn,showpacs,preprintnumbers,amsmath,amssymb, superscriptaddress]{revtex4-1}
\usepackage{graphicx}% Include figure files
\usepackage{dcolumn}% Align table columns on decimal point
\usepackage{bm}% bold math
\usepackage{fancyhdr}
\usepackage{float}
\usepackage{ifthen}
\usepackage{eurosym}
\usepackage{wrapfig}
\usepackage{xfrac}
\usepackage{color}
\usepackage{physics}

\usepackage[normalem]{ ulem }
\usepackage{soul}

\begin{document}
\title{Evolution of Magnetic Order from the Localized to the Itinerant Limit}

\ \author{D.~G.~Mazzone}
\altaffiliation{daniel.mazzone@psi.ch}
\ \affiliation{Laboratory for Neutron Scattering and Imaging, Paul Scherrer Institut, 5232 Villigen PSI, Switzerland}
\ \affiliation{National Synchrotron Light Source II, Brookhaven National Laboratory, Upton, New York 11973, USA}

\ \author{N.~Gauthier}
\altaffiliation{nicolas.gauthier4@gmail.com}
\ \affiliation{Laboratory for Scientific Developments and Novel Materials, Paul Scherrer Institut, 5232 Villigen PSI, Switzerland}
\ \affiliation{Stanford Institute for Materials and Energy Sciences, SLAC National Accelerator Laboratory, Menlo Park, California 94025, USA.}

\ \author{D. T.~Maimone}
\ \affiliation{Laboratory for Scientific Developments and Novel Materials, Paul Scherrer Institut, 5232 Villigen PSI, Switzerland}

\ \author{R.~Yadav}
\ \affiliation{Laboratory for Scientific Developments and Novel Materials, Paul Scherrer Institut, 5232 Villigen PSI, Switzerland}

\ \author{M.~Bartkowiak}
\ \affiliation{Laboratory for Scientific Developments and Novel Materials, Paul Scherrer Institut, 5232 Villigen PSI, Switzerland}

\ \author{J.~L.~Gavilano}
\ \affiliation{Laboratory for Neutron Scattering and Imaging, Paul Scherrer Institut, 5232 Villigen PSI, Switzerland}

\ \author{S.~Raymond}
\ \affiliation{Univ. Grenoble Alpes, CEA, IRIG, MEM, MDN, F-38000 Grenoble, France}

\ \author{V.~Pomjakushin}
\ \affiliation{Laboratory for Neutron Scattering and Imaging, Paul Scherrer Institut, 5232 Villigen PSI, Switzerland}

 \author{N. Casati}
\ \affiliation{Swiss Light Source, Paul Scherrer Institut, CH-5232 Villigen PSI, Switzerland}

 \author{Z. Revay}
\ \affiliation{Technische Universit\"at M\"unchen, Heinz Maier-Leibnitz Zentrum, 85747 Garching, Germany}

\ \author{G.~Lapertot}
\ \affiliation{Univ. Grenoble Alpes, CEA, IRIG, PHELIQS, IMAPEC, F-38000 Grenoble, France}

\ \author{R.~Sibille}
\ \affiliation{Laboratory for Neutron Scattering and Imaging, Paul Scherrer Institut, 5232 Villigen PSI, Switzerland}

\ \author{M.~Kenzelmann}
\ \affiliation{Laboratory for Neutron Scattering and Imaging, Paul Scherrer Institut, 5232 Villigen PSI, Switzerland}

\date{\today}% It is always \today, today,
             
\begin{abstract}
Quantum materials that feature magnetic long-range order often reveal complex phase diagrams when localized electrons become mobile. In many materials magnetism is rapidly suppressed as electronic charges dissolve into the conduction band. In materials where magnetism persists, it is unclear how the magnetic properties are affected. Here we study the evolution of the magnetic structure in Nd$_{1-x}$Ce$_x$CoIn$_5$ from the localized to the highly itinerant limit. We observe two magnetic ground states inside a heavy-fermion phase that are detached from unconventional superconductivity. The presence of two different magnetic phases provides evidence that increasing charge delocalization affects the magnetic interactions via anisotropic band hybridization.
\end{abstract}

\maketitle

Charge carriers in a periodic array of atoms are found either close to the nuclei or assume a delocalized character where they move freely throughout the crystal. These extreme cases are often well-described in the Mott, Kondo or Fermi liquid theory framework. Materials in intermediate regimes with strong electronic fluctuations are more difficult to describe and can stabilize novel quantum ground states that keep fascinating the condensed matter physics community.  Examples include the transition-metal oxides, pnictides or the heavy-fermion metals \cite{DoironLeyraud2012, Kordyuk2012, Mathur1998, Catalano2018}.

In materials with partly delocalized electrons, orbital and spin degrees of freedom can trigger magnetic long-range order. Magnetism is often suppressed by increasing charge delocalization, and it is thought that the associated fluctuations are essential for macroscopically coherent phases such as unconventional superconductivity \cite{DoironLeyraud2012, Kordyuk2012, Mathur1998}. It is not clear how the magnetic interactions, and the resulting magnetic structure, are modified as the degree of itineracy changes in a material. Ab-initio calculations are at present not accurate enough to deal with the small energy differences that are relevant, particularly in heavy-fermion metals. In addition, it is an open question how the magnetic interactions are affected in experimental realizations where magnetic order persists over localized-to-itinerant charge transitions \cite{Friedemann2009, Jiao2014, Custers2012}.

%Ab-initio calculations are at present not accurate enough to deal with the small energy differences that are relevant in heavy-fermion metals, in particular. In addition, in experimental realizations where magnetic order persists over localized-to-itinerant charge transitions, it is not clear how the magnetic interactions are affected \cite{Friedemann2009, Jiao2014, Custers2012}.

\begin{figure}[tbh]
\includegraphics[width=\linewidth]{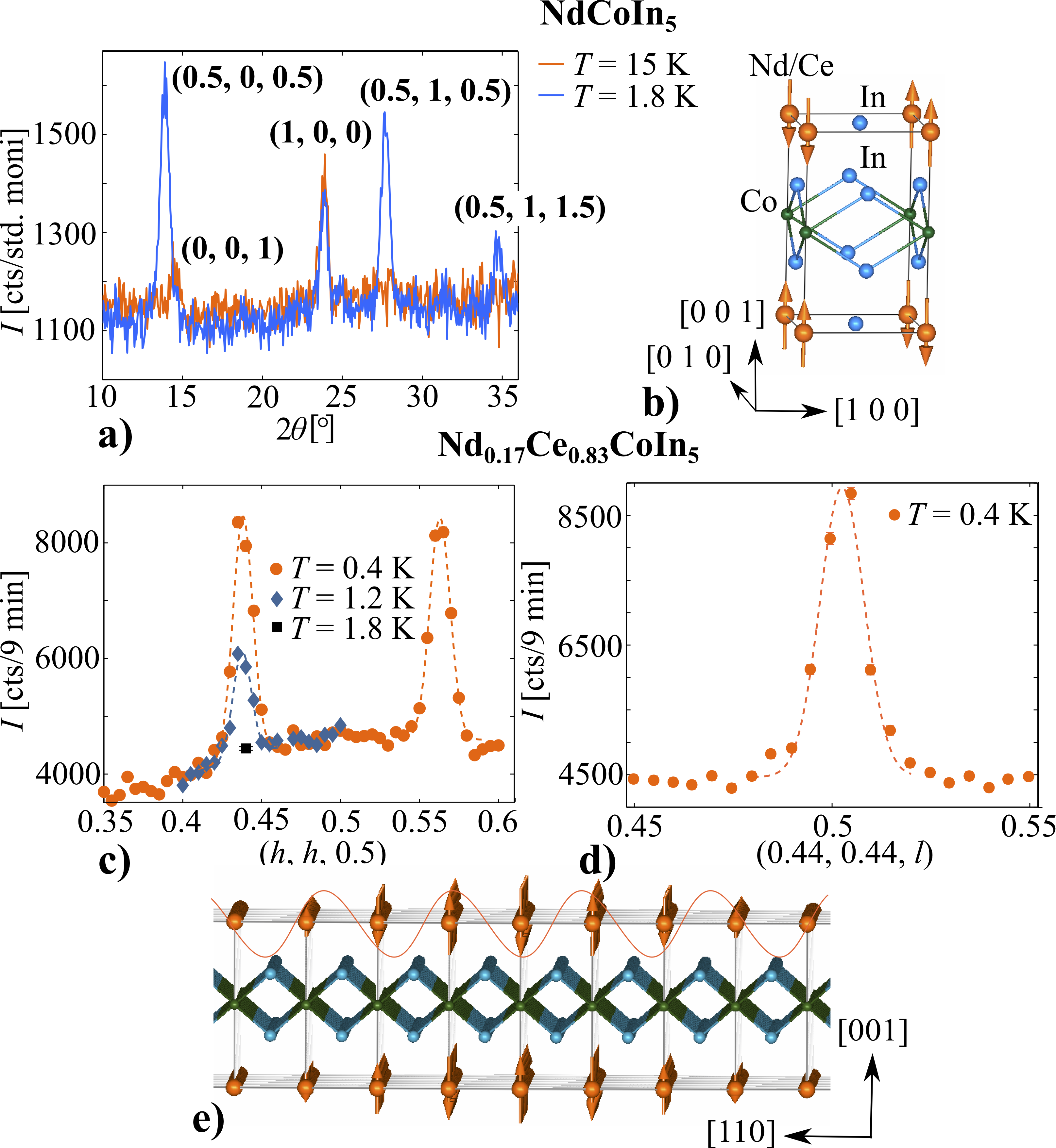}
\caption{(Color online) \textbf{a)} Neutron powder diffraction results on NdCoIn$_5$ revealing an Ising-like magnetic structure, as shown in  \textbf{b)}. Single crystal neutron diffraction intensity of Nd$_{0.17}$Ce$_{0.83}$CoIn$_5$ along ($h$, $h$, 0.5) in  \textbf{c)} and (0.44, 0.44, $l$) in  \textbf{d)}, given in reciprocal lattice units (r.l.u). \textbf{e)} Amplitude modulated magnetic structure, such as observed for 0.75 $\leq$ x $\leq$ 0.95.}
\label{fig1}
\end{figure}

Kondo materials are ideal systems to shed light onto this scientific problem. The materials possess energy scales that are up to four orders of magnitude smaller than in  transition-metal oxides \cite{Aynajian2012}, and are, thus, highly susceptible to external tuning parameters, such as magnetic fields or chemical doping \cite{Weng2016, Wirth2016, Dzero2016}. The low energy scale arises from partially-filled electronic $f$-states that are partly screened by mobile charge carriers. This so-called Kondo coupling can lead to collective singlet states, where localized $f$-electrons admix with the conduction band and become mobile. The development of coherent $f$-bands near the Fermi surface triggers renormalized effective masses below a coherence temperature $T_{coh}$ that allows to effectively probe the degree of the system's itinerancy. In addition, heavy-fermion materials can feature localized magnetic order that is in direct competition with Kondo screening, providing an accessible and sensitive measure of the nature of magnetism. 

Here, we study the evolution of magnetic properties in Nd$_{1-x}$Ce$_x$CoIn$_5$. CeCoIn$_5$ is a highly itinerant heavy-fermion material with a large quasiparticle mass enhancement \cite{Petrovic2001, Nakatsuji2004}. In contrast, isotroctural compounds consisting of Nd ions are known to generate non-hybridized local magnetic moments \cite{Cermak2014, Chang2002, Hieu2006, Hu2008}. Previous macroscopic transport measurements have shown that chemical substitution of Nd for Ce in Nd$_{1-x}$Ce$_x$CoIn$_5$ allows to drive the system from the highly itinerant to the completely localized limit \cite{Hu2008}. The series displays zero-field magnetic order for $x$ $<$ 0.95 that competes with superconductivity for $x$ $\geq$ 0.83, and heavy-electron bands are thought to exist for $x$ $>$ 0.5. Hitherto, magnetic order has been explored in detail for $x$ = 0.95 only, where a spin-density wave (SDW) is modulated with $\vec{Q}_{IC}$ = ($q$, $\pm q$, 0.5) in reciprocal lattice units (r.l.u) and $q$ $\approx$ 0.45 \cite{Raymond2014, Mazzone2017}. The ordered magnetic moment, $\mu$ = 0.13(5)$\mu_B$, is aligned along the tetragonal $c$-axis. As we will now show, the magnetic symmetry is modified within the heavy-fermion ground state upon doping, providing evidence for an evolution of the magnetic exchange couplings upon band hybridization. 

Single crystalline Nd$_{1-x}$Ce$_x$CoIn$_5$ with $x$ = 0, 0.16, 0.4, 0.61, 0.75, 0.83, 0.95 and 1 was grown in indium self-flux \cite{Petrovic2001}. The quality of experimental realization with $x$ = 0, 0.16, 0.4, 0.61 and 1 was probed via X-ray powder diffraction at the Material Science (MS-X04SA) beamline of the Swiss Light Source at the Paul Scherrer Institut (PSI), Villigen, Switzerland using a photon wave-length $\lambda$ = 0.56491 \AA~\cite{Willmott2013}. The actual Nd concentration of these samples was determined via high-resolution neutron diffraction on HRPT at the Swiss Neutron Spallation Source (SINQ) at PSI with  $\lambda$ = 1.886 \AA. The Nd concentration in single crystals with $x$ = 0.95 was checked by means of in-beam neutron activation analysis at MLZ-Garching, Munich, Germany \cite{Revay2015, Revay2009}. The macroscopic properties of members with $x$ = 0, 0.16, 0.4, 0.61, 0.75, 0.83 and 1 were investigated via four-probe electrical resistivity and AC/DC magnetization measurements in a Quantum Design PPMS or in cryogenic magnets. The macroscopic and microscopic properties of $x$ = 0.95 are reported in Ref. 19. The magnetic structure of $x$ = 0, 0.16, 0.4, 0.61 was determined via neutron powder diffraction at HRPT and experimental realizations with $x$ = 0.75 and 0.83 were investigated by means of single crystal neutron diffraction on Zebra at SINQ and on the triple-axis spectrometer IN12 at the Institut Laue-Langevin, Grenoble France, respectively. While a neutron wave-length $\lambda$ = 1.177 \AA~was employed on Zebra, $\lambda$ = 3.307 and 4.83 \AA~were used on IN12. All magnetic structures were refined with the FullProf suite \cite{Fullprof}. Further detailed information are given in the Supplemental Material.

Neutron powder diffraction results on NdCoIn$_5$ are shown in Fig. \ref{fig1}a and are representative for Nd$_{1-x}$Ce$_x$CoIn$_5$ with $x$ = 0, 0.16, 0.4 and 0.61 (see Supplemental Material). We observe magnetic Bragg peaks at low temperatures, consistent with two symmetry-equivalent commensurate propagation vectors $\vec{Q}_{C}$ = (1/2, 0, 1/2) and (0, 1/2, 1/2). They correspond to different domains that are indistinguishable in a powder diffraction experiment. The evolution of the magnetic Bragg peak intensity at higher scattering angles shows unambiguously that the magnetic moment is oriented along the $c$-axis. We find an Ising-like structure that is displayed in Fig. \ref{fig1}b for the (1/2, 0, 1/2)-domain.
\begin{figure}[tbh]
\includegraphics[width=\linewidth]{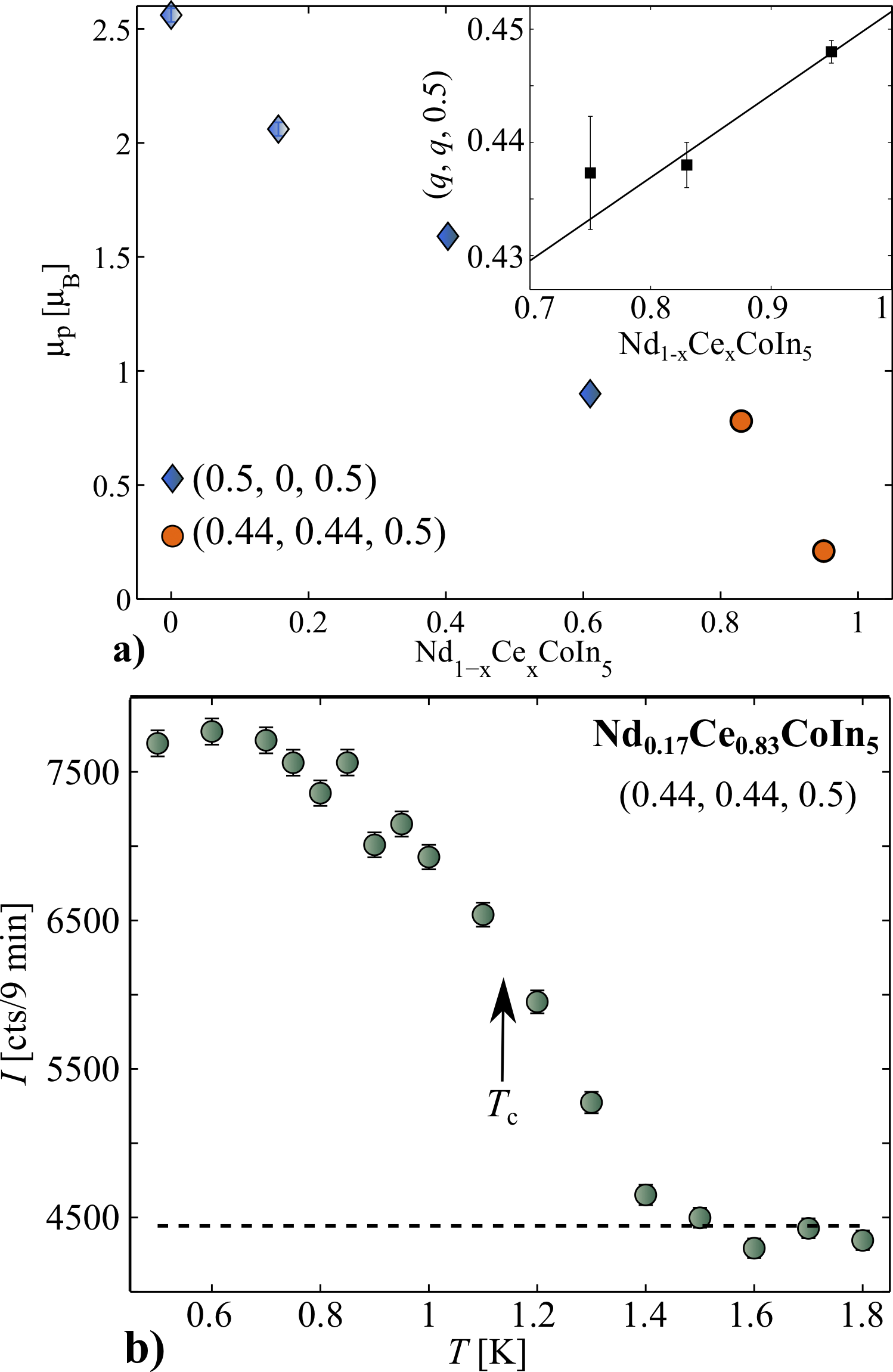}
\caption{(Color online) \textbf{a)} Evolution of the magnetic moment amplitude as a function of Ce content in Nd$_{1-x}$Ce$_x$CoIn$_5$ (the data point at $x$ = 0.95 was taken from Ref. 19). Inset: Incommensuration ($q$, $q$, 0.5) as a function of $x$. \textbf{b)} Temperature dependence of the magnetic Bragg peak intensity at (0.44, 0.44, 0.5) in Nd$_{0.17}$Ce$_{0.83}$CoIn$_5$.}
\label{fig2}
\end{figure}

In contrast, a ground state with incommensurate magnetic order is observed for $x$ $\geq$ 0.75. The single crystal neutron diffraction results reveal a magnetic wave-vector $\vec{Q}_{IC}$ = ($q$, $\pm q$, 0.5) with $q$ $\approx$ 0.44,  for Nd$_{1-x}$Ce$_x$CoIn$_5$ with 0.75 $\leq$ $x$ $\leq$ 0.95 (see Fig. \ref{fig1}c and d for $x$ = 0.83 and Supplemental Material for $x$ = 0.75). The propagation vector is similar to the one of Nd$_{0.05}$Ce$_{0.95}$CoIn$_5$ \cite{Raymond2014, Mazzone2017}, and the absence of intensity at $\vec{Q}_{C}$ excludes a scenario where different types of magnetic order coexist. The magnetic moment remains aligned along the tetragonal $c$-axis, but features a sinusoidally modulated structure (see Fig. \ref{fig1}e).

\begin{figure}[tbh]
\includegraphics[width=\linewidth]{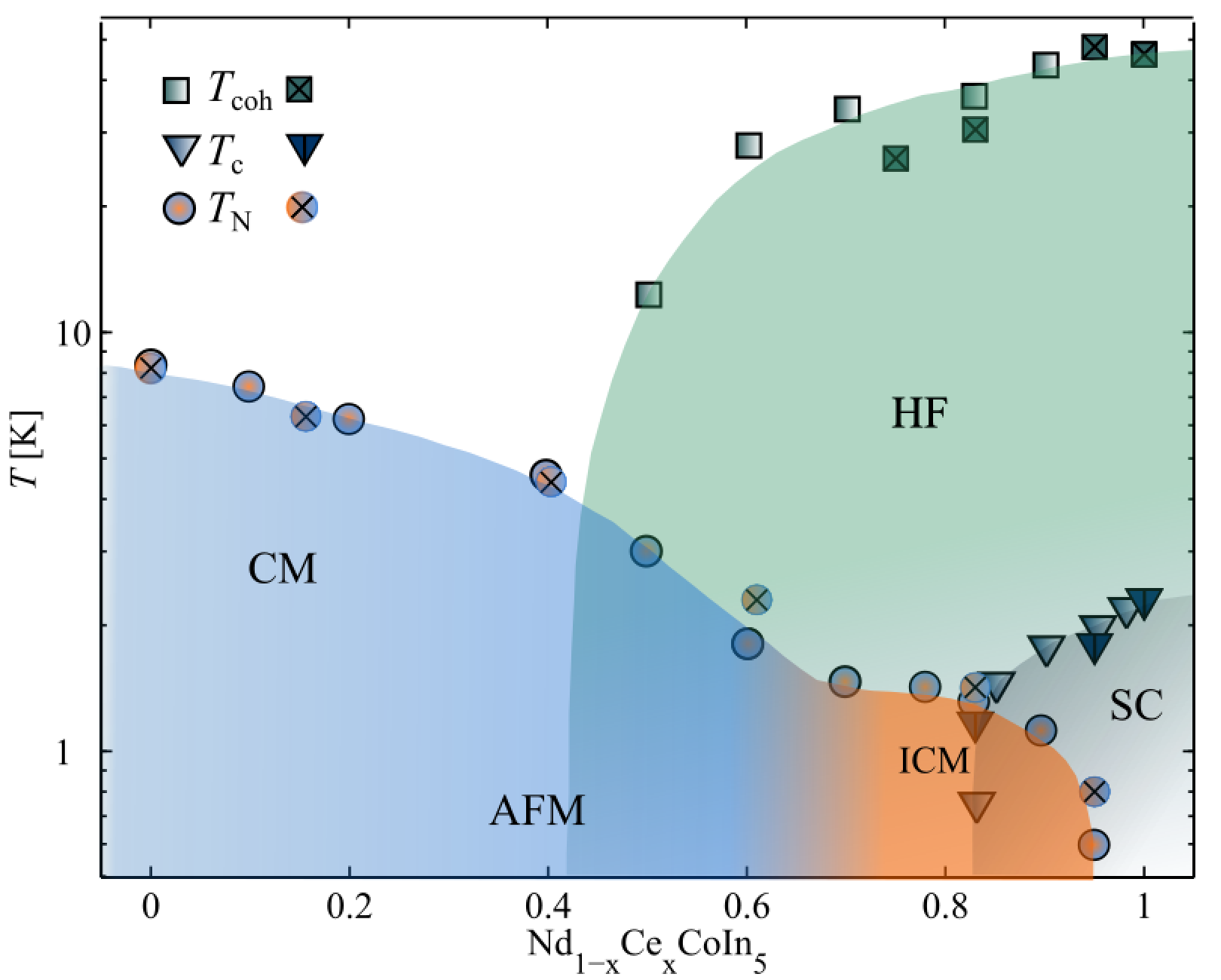}
\caption{(Color online)  Antiferromagnetic (AFM) phase for $x$ $\leq$ 0.95, superconductivity (SC) for $x$ $\geq$ 0.83 and heavy-fermion (HF) properties for $x$ $\geq$  0.5. CM denotes the commensurate structure with $\vec{Q}_{C}$ = (1/2, 0, 1/2) and (0, 1/2, 1/2). ICM is the incommensurate order along $\vec{Q}_{IC}$ = ($q$, $\pm q$, 0.5) with $q$ $\approx$ 0.44. Left legend represent data from Ref. 17, the right legend are our data (the phase boundaries of $x$ = 0.95 were taken from Ref. 19).}
\label{fig3}
\end{figure}

The doping dependence of the magnetic moment amplitude is shown in Fig. \ref{fig2}a. In the commensurate phase, it monotonically decreases from $\mu_p$ = 2.56(3)$\mu_B$ for NdCoIn$_5$ to 0.90(5)$\mu_B$ for a Ce concentration of 61\%. The modulation direction experiences a rotation in the tetragonal plane and propagates along $\vec{Q}_{IC}$ for $x$ $\geq$ 0.75. Here $\mu_p$ is modified weakly before it is strongly suppressed for $x$ $>$ 0.83. The incommensuration, $q$($x$), increases with increasing Ce content (see Fig. \ref{fig2}a inset), which may be attributed to small changes in the Fermi surface. The $xT$-phase diagram of Nd$_{1-x}$Ce$_x$CoIn$_5$ is shown in Fig. \ref{fig3}. The series features persistent magnetism up to $x$ $=$ 0.95 that competes with superconductivity at $x$ $\geq$ 0.83, and a localized charge state for $x$ $<$ 0.5. The key result of our study is that magnetic order is modified between $x$ = 0.61 and 0.75, shifted with respect to the onset of coherent heavy bands and superconductivity.

The superconducting phase at $x$ $>$ 0.8 is believed to arise from magnetic fluctuations of a  nearby SDW critical point \cite{Tokiwa2013}. The Cooper-pairs in CeCoIn$_5$ feature d$_{x^2-y^2}$-symmetry that is robust under small Nd substitution \cite{Allan2013, Kim2015}. The pairing symmetry reveals nodes along the reciprocal (1, 1, 0)-direction, where low-energy quasiparticles can mediate magnetic order without directly competing with the condensate. In consequence, the $d$-wave order parameter is compatible with the incommensurate wave-vector $\vec{Q}_{IC}$, but not with $\vec{Q}_{C}$. This is supported by the temperature dependence of the magnetic Bragg peak intensity in Nd$_{0.17}$Ce$_{0.83}$CoIn$_5$ that shows no anomaly as the temperature is tuned across the superconducting phase boundary (see Fig. \ref{fig2}b).

The interplay between superconductivity and magnetism observed here is very different from the behavior in isostructural CeCo$_y$Rh$_{1-y}$In$_5$, where the magnetic moment orientation is altered at the superconducting phase boundary \cite{Kim2009, Kawamura2007, Yokoyama2008}. In Nd$_{1-x}$Ce$_x$CoIn$_5$ magnetic order along $\vec{Q}_{C}$ is established between 0.61 $<$ $x$ $<$ 0.75 where superconductivity is suppressed (see Fig. \ref{fig3}). In contrast, similarities with the $xT$-phase digram of the iron-based superconductor Fe$_{1+y}$Te$_{1-z}$Se$_z$ are recognized \cite{Liu2010}. The material hosts two different types of antiferromagnetic correlations, whose relative weight can be tuned via Se substitution. The dominant correlations at low Se concentrations are associated to weak charge carrier localization that triggers magnetic long-range order. In contrast, antiferromagnetic correlations with a different modulation become important at larger Se concentrations and are thought to be closely related to emergence of superconductivity. Similarly, inelastic neutron scattering studies on Nd$_{1-x}$Ce$_x$CoIn$_5$ with $x$ = 1 and 0.95 have shown that the magnetic fluctuations related to superconductivity possess a symmetry distinct from the magnetic modulation in the localized limit \cite{Raymond2012, Mazzone20172}. Thus, an intimate link between magnetic correlations along $\vec{Q}_{IC}$ and d$_{x^2-y^2}$-wave superconductivity is expected in Nd$_{1-x}$Ce$_x$CoIn$_5$.

%the symmetry of magnetic correlations and the emergence of superconductivity is also expected in this series.
%A change in the magnetic structure at the superconducting phase boundary is found in isostructural CeCo$_y$Rh$_{1-y}$In$_5$ \cite{Raymond2008, Kim2009}. In strong contrast, the magnetic phase transition in Nd$_{1-x}$Ce$_x$CoIn$_5$ is not directly connected to the superconducting ground state, as it arises for Ce concentrations 0.61 $<$ $x$ $<$ 0.75 where superconductivity is already suppressed (see Fig. \ref{fig3}). 

%This behavior shares some similarities with the $xT$-phase digram of the iron-based superconductor Fe$_{1+y}$Te$_{1-z}$Se$_z$ \cite{Liu2010}. The material hosts two different types of antiferromagnetic correlations, whose relative weight can be tuned via Se substitution. The dominant correlations at low Se concentrations are associated to weak charge carrier localization that triggers magnetic long-range order. In contrast, antiferromagnetic correlations with a different modulation become important at larger Se concentrations and are thought to be closely related to emergence of superconductivity. Similarly, inelastic neutron scattering studies on Nd$_{1-x}$Ce$_x$CoIn$_5$ with $x$ = 1 and 0.95 have shown that the magnetic fluctuations related to superconductivity posses a symmetry distinct from the magnetic modulation in the localized limit \cite{Raymond2012, Mazzone20172}. Thus, an intimate link between the symmetry of magnetic correlations and the emergence of superconductivity is also expected in this series.  

\begin{figure}[tbh]
\includegraphics[width=\linewidth]{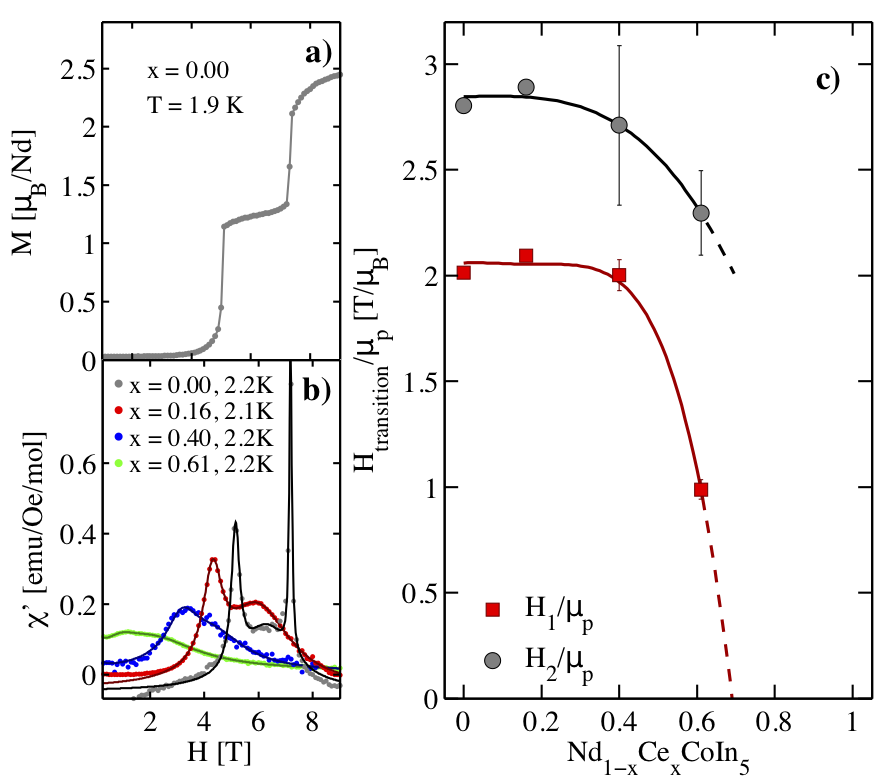}
\caption{(Color online) \textbf{a)} Field dependent DC magnetization, $M$, of NdCoIn$_5$ and \textbf{b)} real part of the AC magnetic susceptibility, $\chi'$ in Nd$_{1-x}$Ce$_x$CoIn$_5$ with $x$ = 0, 0.16, 0.4 and 0.61 for $\vec{H}||$[0 0 1] and $T$ $<$ $T_N$ . The two transitions correspond to a spin-flip ($H_1$) and a ferromagnetic transition ($H_2$).  \textbf{c)} Field dependence of the two critical fields, $H_{1,2}$, normalized by the ordered moment, $\mu_p$. The solid lines are guides to the eye.}
\label{fig4}
\end{figure}
A change of magnetic correlations points towards a doping dependent evolution of the magnetic interactions that is related to the degree of electron mobility. This is observed, for instance, in some layered transition-metal-oxides, as they establish a charge- and spin-stripe order that allows to accommodate coexisting local antiferromagnetic spin correlations and itinerant charge carriers \cite{Tranquada2012}. In La$_{2-y}$Ba$_y$CuO$_4$, for instance, the antiferromagnetic Mott-ground state is suppressed for $y$ $\geq$ 0.05, but the inclusion of mobile hole carriers affects the magnetic exchange couplings stabilizing a stripe order. This ground state competes with superconductivity that is also present for $y$ $\geq$ 0.05, causing a profound reduction of the critical temperature around $y$ $\approx$ 1/8. Similarly, the magnetic interactions in Nd$_{1-x}$Ce$_x$CoIn$_5$ may evolve with increasing Ce concentration, and trigger a transition from localized $\vec{Q}_{C}$ to itinerant $\vec{Q}_{IC}$.

The scenario is further clarified by a comparison with isostructural localized-moment magnets that also show a commensurate structure along $\vec{Q}_{C}$ \cite{Cermak2014, Chang2002, Hieu2006}. The materials feature magnetic order that can be described using an Ising-type Hamiltonian with antiferromagnetic nearest-, next-nearest-neighbor and an inter-layer exchange coupling \cite{Hieu2006}. Field dependent magnetization and susceptibility measurements inside the antiferromagnetic phase for fields along the tetragonal $c$-axis can reveal valuable information about the exchange couplings. They show two critical fields, $H_{1,2}$, and a plateau region with increasing magnetic field. The ratio of $H_{1,2}$ and the ordered moment, $\mu_p$, is directly related to the magnetic exchange couplings \cite{Hieu2006}. Similar measurements on Nd$_{1-x}$Ce$_x$CoIn$_5$ with $x$ = 0, 0.16, 0.4 and 0.6 are shown in Fig. \ref{fig4}a and b and the doping dependence of $H_{1,2}$/$\mu_p$ is displayed in Fig. \ref{fig4}c. These results reveal a significant change of the magnetic interactions for $x$ $>$ 0.5, coinciding with the onset of the heavy-fermion ground state where localized $f$-electrons become mobile via hybridization with the conduction band. 

% A change of the magnetic interactions upon doping is supported by field dependent magnetization and susceptibility measurements inside the antiferromagnetic phase for fields along the tetragonal $c$-axis (see Fig. \ref{fig4}a and b). The results show two critical fields, $H_{1,2}$, and a plateau region with increasing magnetic field. The ratio of $H_{1,2}$ and the ordered moment, $\mu_p$, is directly related to the magnetic exchange couplings \cite{Hieu2006}. Its doping dependence is shown in Fig. \ref{fig4}c and reveals a significant change of the magnetic interactions for $x$ $>$ 0.5, coinciding with the onset of the heavy-fermion ground state. Here localized $f$-electrons become mobile via hybridization with the conduction band. 

The underlying microscopic process of the heavy-fermion ground state is driven through collective screening of the conduction electrons, minimizing the total angular momentum of the local 4$f$-moments. While this is favorable for Ce-ions (Ce$^{3+}$,  $J$ = 5/2 $\rightarrow$ Ce$^{4+}$, $J$ = 0), Nd$^{3+}$ remains in a stable $J$ = 9/2 configuration. Such a case can be described best using a phenomenological two-fluid model, in which two coexisting contributions of either $f$-electrons that are hybridized at low temperatures or residual local moments are assumed \cite{Nakatsuji2004, Yang2014, Yang2016}. Since the Ce-4$f$ ground state wave-function in this family of compounds is hybridized mainly with In-$p$ orbitals \cite{Haule2010, Willers2010, Willers2015}, it is conceivable that the antiferromagnetic nearest-, next-nearest-neighbor and inter-layer exchange coupling are affected differently as band hybridization becomes stronger with increasing Ce concentration  (see Fig. \ref{fig1}b). This scenario can naturally explain a change in magnetic symmetry inside the heavy-fermion ground state, as opposed to a case where all interactions change on an equal footing. 

It is noted that the evolution of the hybridized 4$f$-wave function in doped CeCoIn$_5$ strongly depends on the chemical element that is substituted. While Ce-substitution simply yields a decrease in the fluid accounting for hybridized $f$-electrons, doping at the other chemical sites have a more complex impact on the electronic ground state \cite{Haule2010, Willers2010, Willers2015, Chen2018}. It has been shown, for instance, that substitution of the transition metal affects the shape of the Ce-4$f$ ground state wave-function \cite{Willers2015}. Upon increasing Rh doping the 4$f$-orbital is squeezed into the tetragonal basal plane, mainly decreasing the hybridization with the out-of-plane In-$p$ orbitals. This drives the system away from the superconducting ground state and into different magnetic phases \cite{Kim2009, Kawamura2007, Yokoyama2008}. Substitution on the In-site can also affect the shape of the Ce-4$f$ ground state wave-function, but a recent study has shown that it remains unchanged under Cd substitution \cite{Chen2018}. In this compound magnetic order is thought to arise via a local nucleation process, without altering the global electronic structure \cite{Sakai2015}.

A similar controversy has also arisen lately for magnetic order in Nd$_{1-x}$Ce$_x$CoIn$_5$ at small $x$ \cite{Martiny2015, Zhu2017, Rosa2017, Green2018}. De Haas-van Alphen effect measurements on Nd$_{1-x}$Ce$_x$CoIn$_5$,  predictions on the behavior of the static spin susceptibility upon (non-)magnetic substitution of Ce in CeCoIn$_5$  and the observation of magnetic order in 5\% Gd-doped CeCoIn$_5$ suggest that magnetic order is triggered by an instability in the band structure \cite{Green2018, Rosa2017}. In contrast, some theoretical models argue that local magnetic droplets play a decisive role \cite{Martiny2015, Zhu2017}. In this scenario magnetism is mediated inside a $d$-wave superconducting background via strong magnetic fluctuations triggered by the nearby SDW critical point. Since we find that magnetic order along $\vec{Q}_{IC}$ persists down to $x$ = 0.75 that is both, outside the superconducting dome and far from the SDW critical point, our results are in line with a non-local picture involving the Fermi surface topology.

In summary, we report the evolution of magnetism in the series Nd$_{1-x}$Ce$_{x}$CoIn$_5$ from the localized ($x$ = 0) to the highly itinerant limit ($x$ = 1). We observe two different magnetic structures with moments along the tetragonal $c$-axis. Magnetic order is Ising-like with $\vec{Q}_{C}$ = (1/2, 0, 1/2) and (0, 1/2, 1/2) for materials with $x$ $\leq$ 0.61, and amplitude modulated with $\vec{Q}_{IC}$ = ($q$, $\pm q$, 0.5) with $q\approx$ 0.44 for $x$ $\geq$ 0.75. We find that delocalization of Ce-4$f$ electrons at $x$ $>$ 0.5 affects the magnetic interactions, providing evidence for anisotropic hybridization effects. Increasing charge itinerancy leads to a magnetic transition between $x$ = 0.61 and 0.75. This occurs far away from the emergence of unconventional superconductivity and is thus unrelated. Our results demonstrate that the magnetic interactions strongly depend on the degree of 4$f$-electron hybridization with the conduction electrons. This emphasizes the need to include hybridization dependent magnetic interaction in theories describing quantum materials close to the Kondo breakdown, and offers new perspectives for the interpretation of the physical properties in heavy-fermion materials.

We thank the Paul Scherrer Institut, the Institut Laue-Langevin and the Forschungsneutronenquelle Heinz Maier-Leibnitz for the allocated beam time. We acknowledge Mark Dean, Maxim Dzero and Priscila Rosa for fruitful discussions, and Maik Locher for his help with the DC magnetization measurements. We thank the Swiss National Foundation (grant No. 200021\_147071, 200021\_138018 and 200021\_157009 and Fellowship No. P2EZP2\_175092 and P2EZP2\_178542) for financial support. This work used resources of the National Synchrotron Light Source II, a U.S. Department of Energy (DOE) Office of Science User Facility operated for the DOE Office of Science by Brookhaven National Laboratory under Contract No. DE-SC0012704.

\def\bibsection{\section*{\refname}}

\begin{figure}[tbh]
\includegraphics[width=\textwidth]{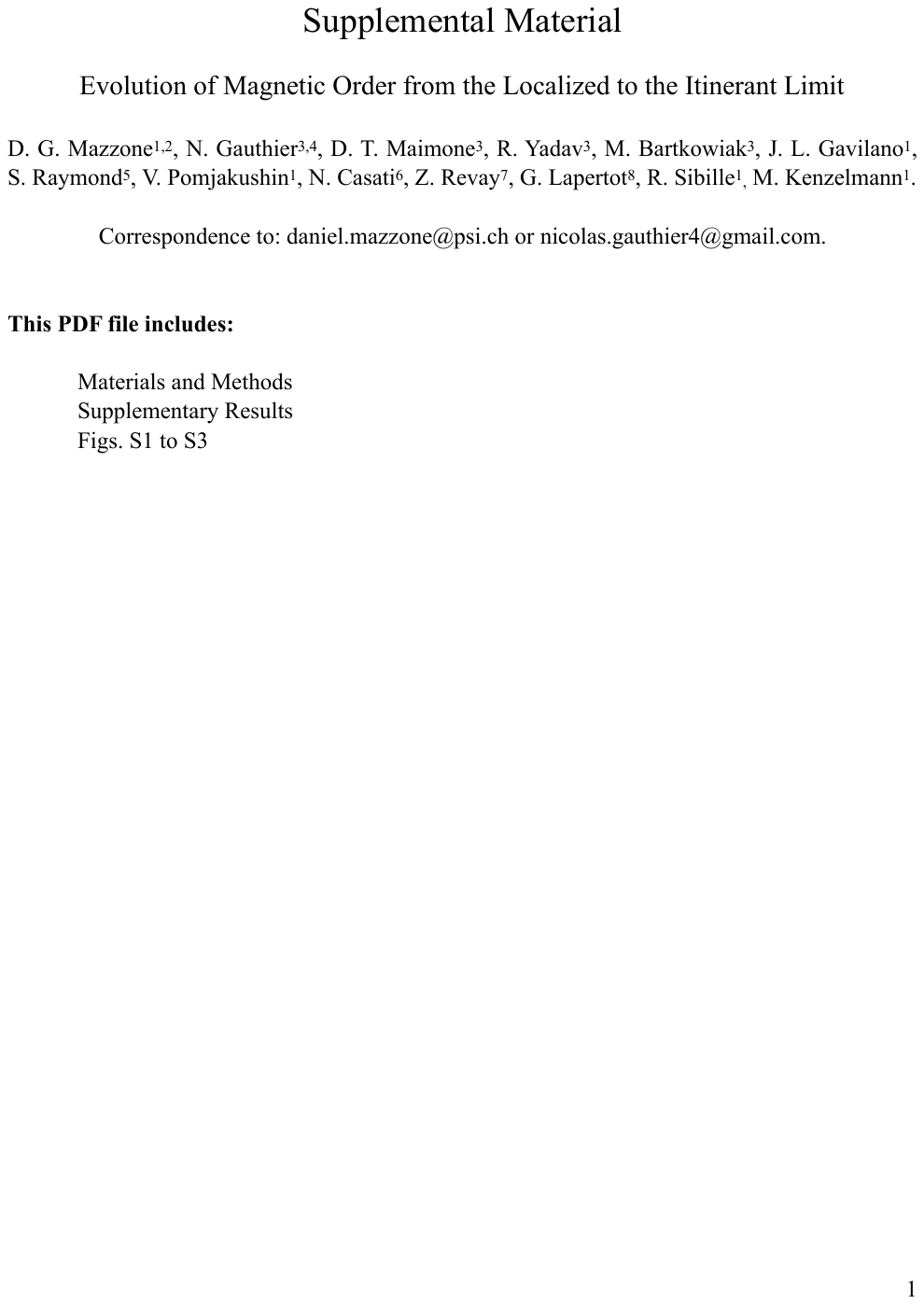}
\end{figure}
\begin{figure}[tbh]
\includegraphics[width=\textwidth]{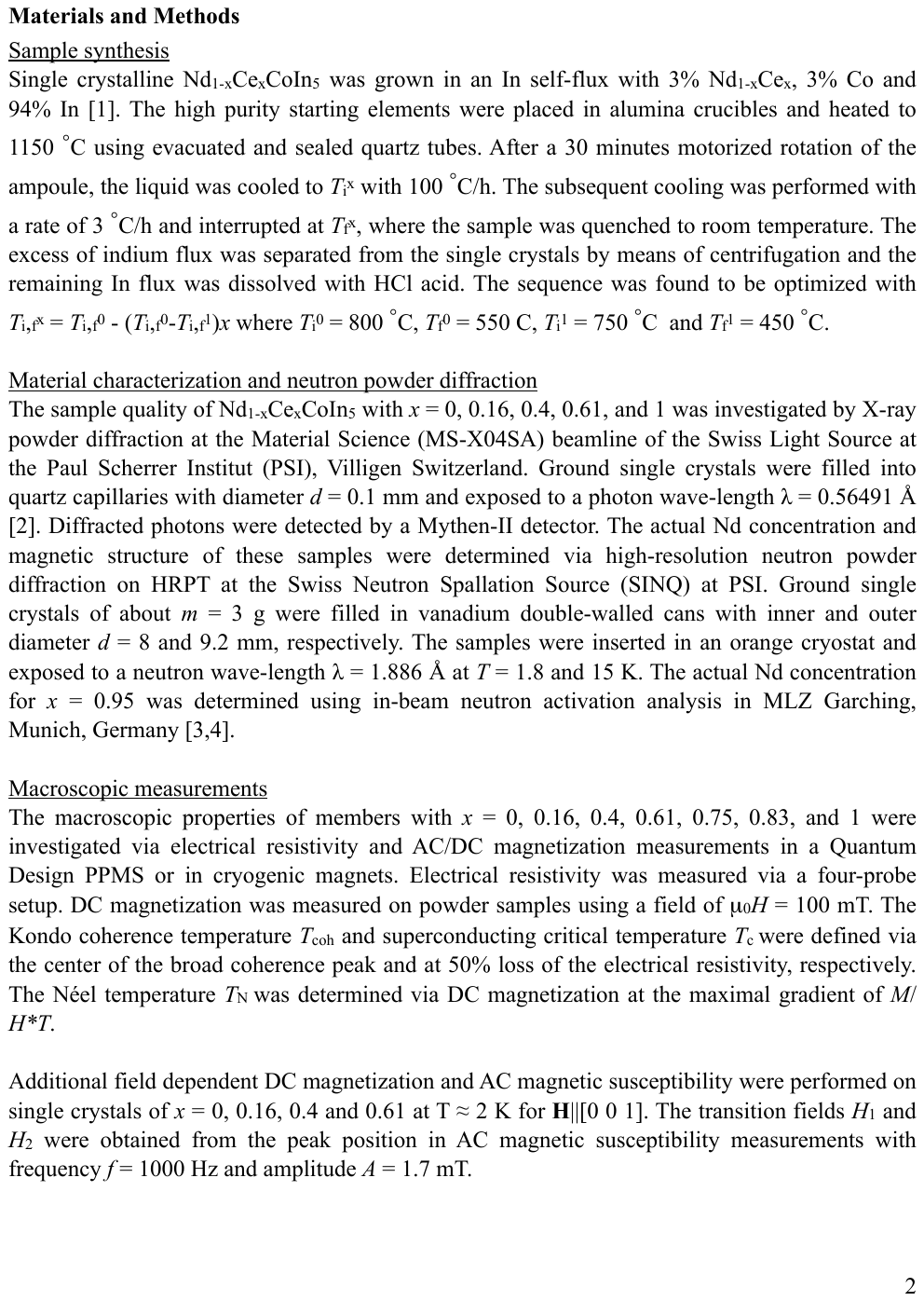}
\end{figure}
\begin{figure}[tbh]
\includegraphics[width=\textwidth]{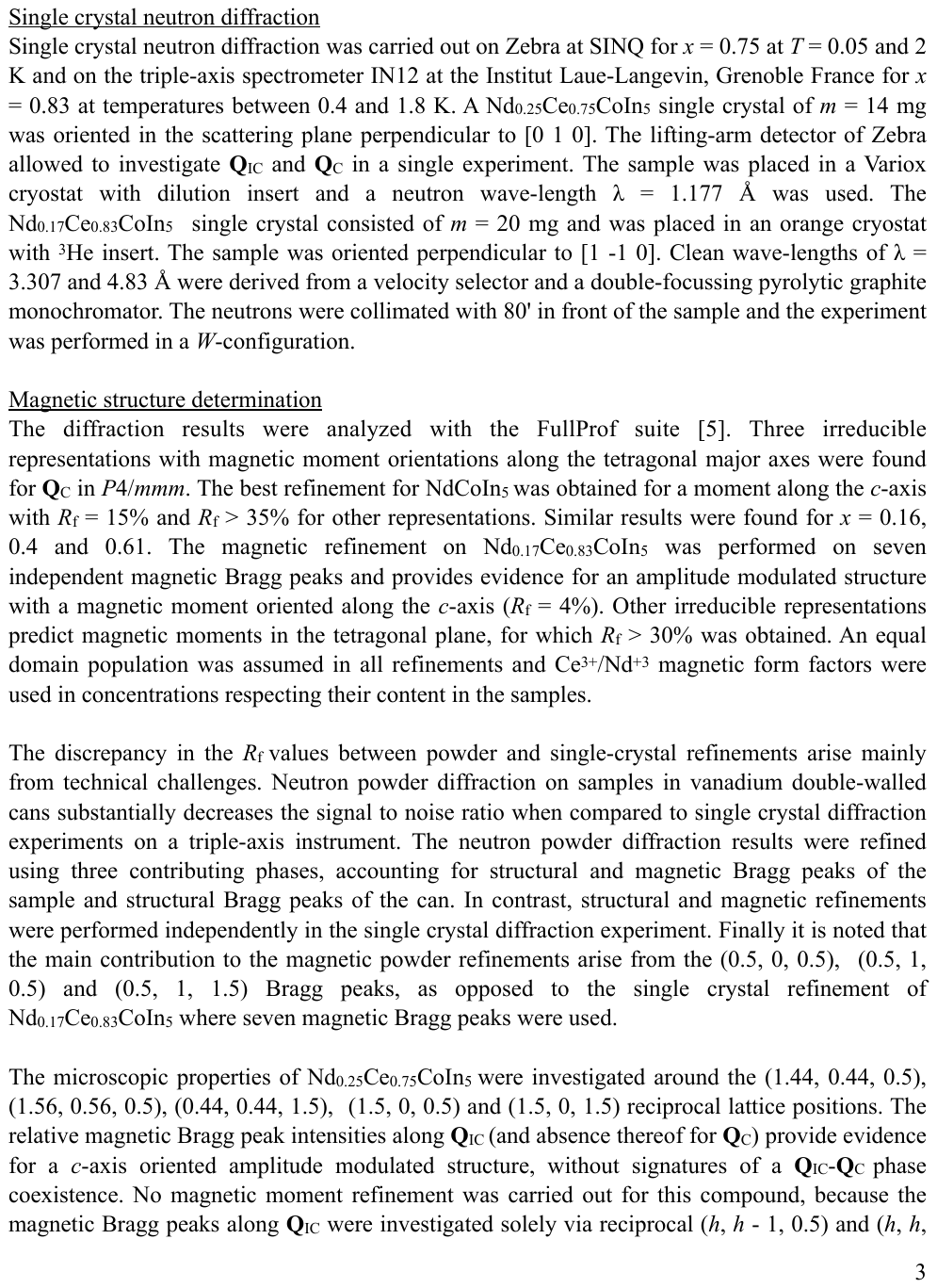}
\end{figure}
\begin{figure}[tbh]
\includegraphics[width=\textwidth]{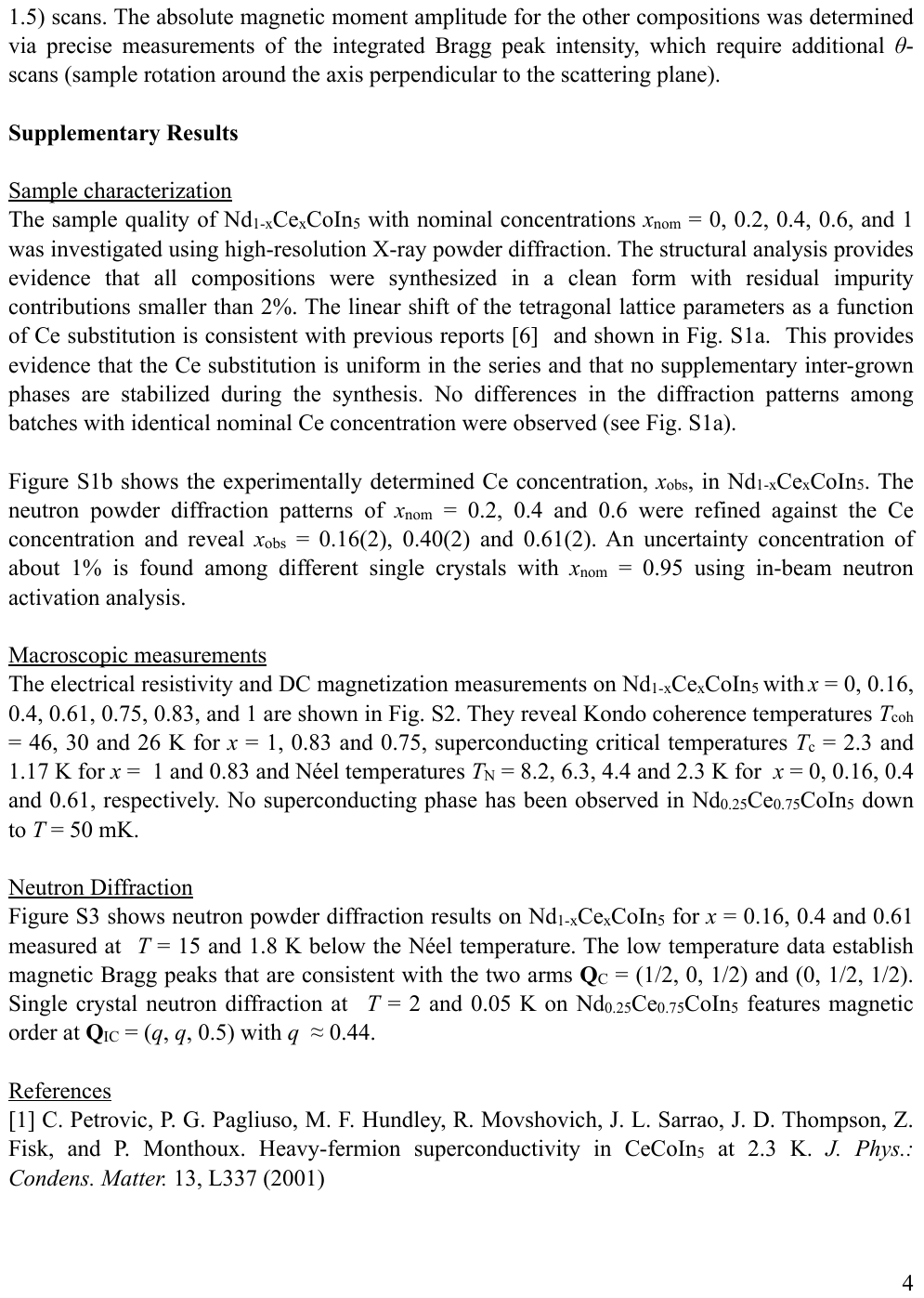}
\end{figure}
\begin{figure}[tbh]
\includegraphics[width=\textwidth]{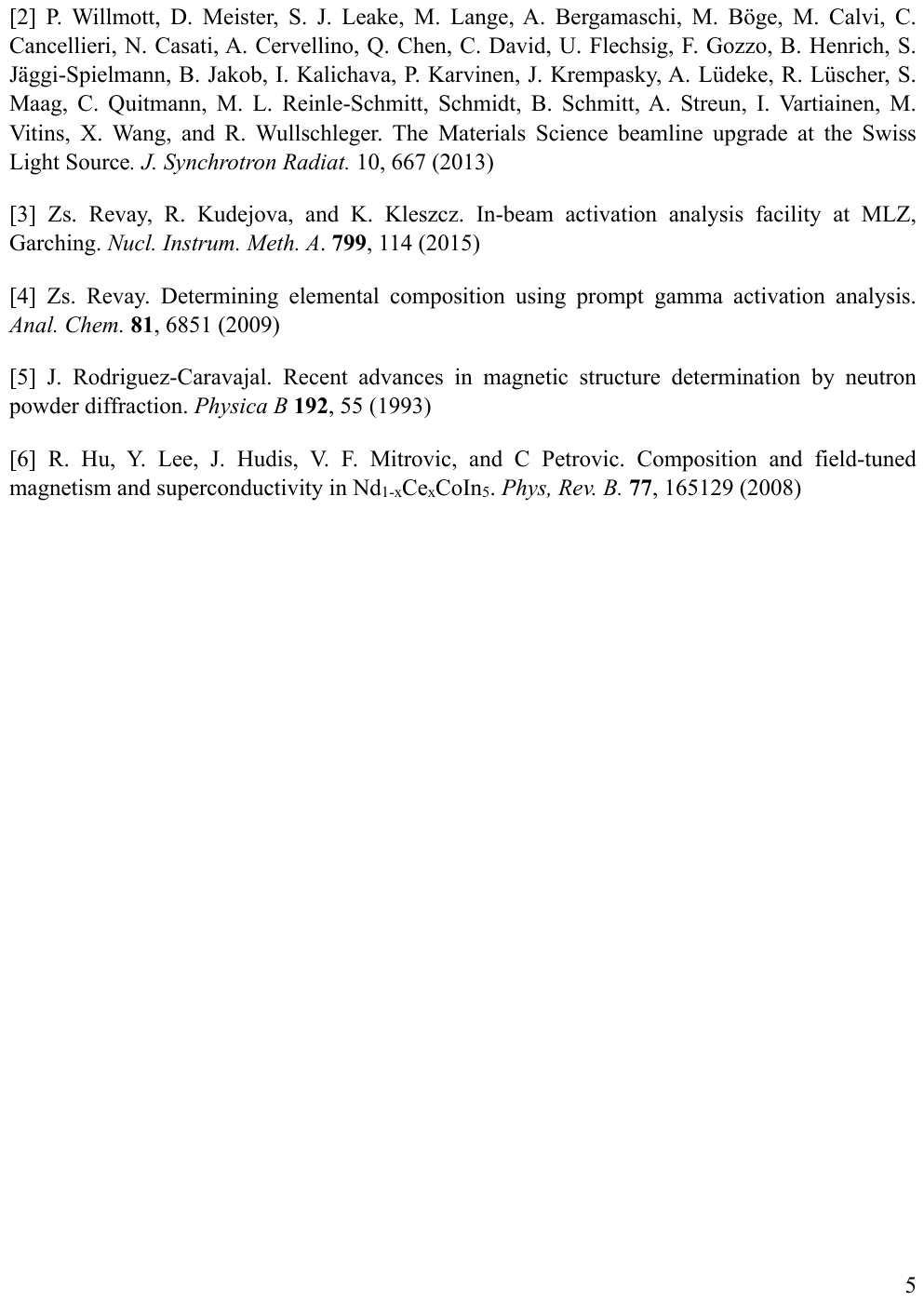}
\end{figure}
\begin{figure}[tbh]
\includegraphics[width=\textwidth]{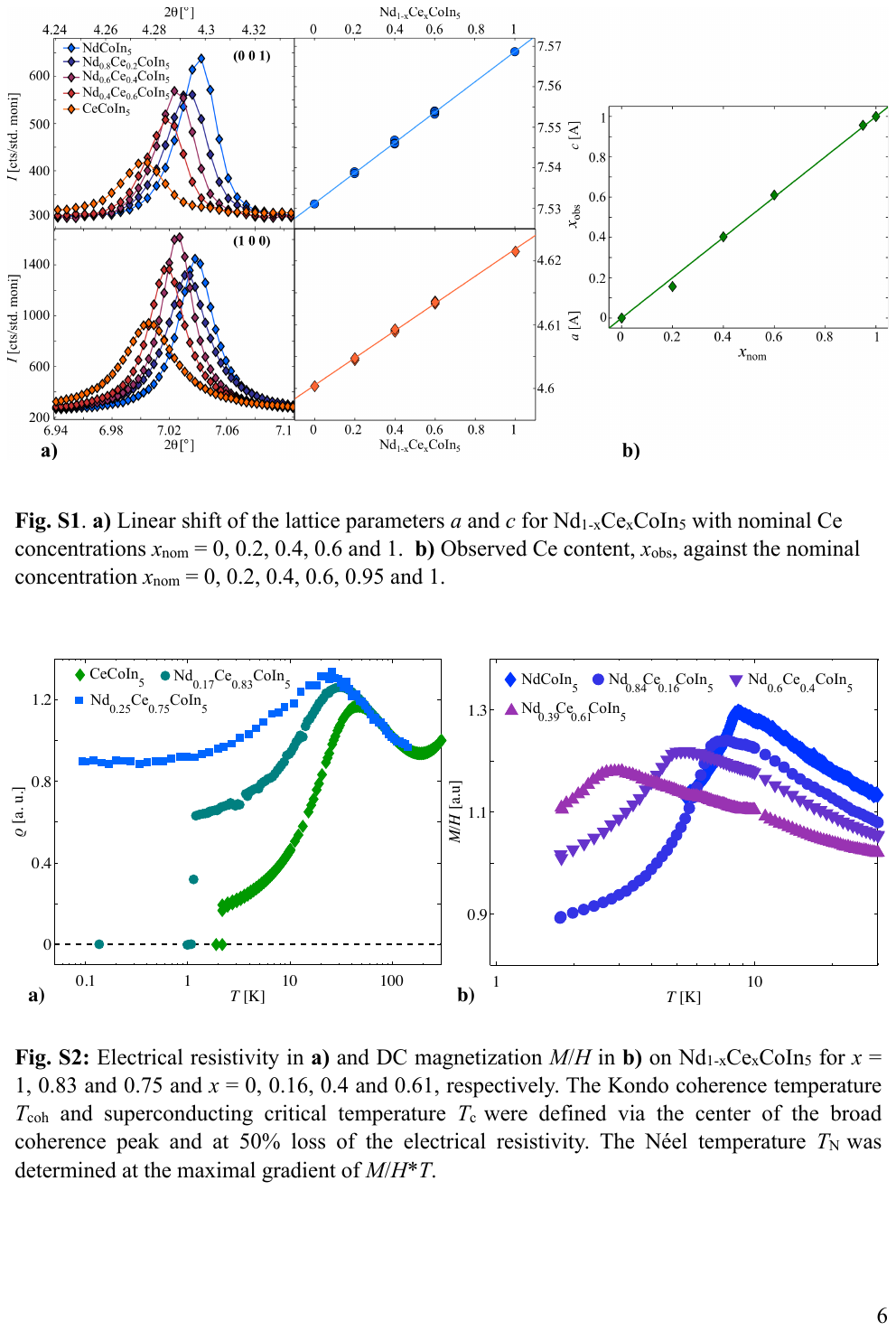}
\end{figure}
\begin{figure}[tbh]
\includegraphics[width=\textwidth]{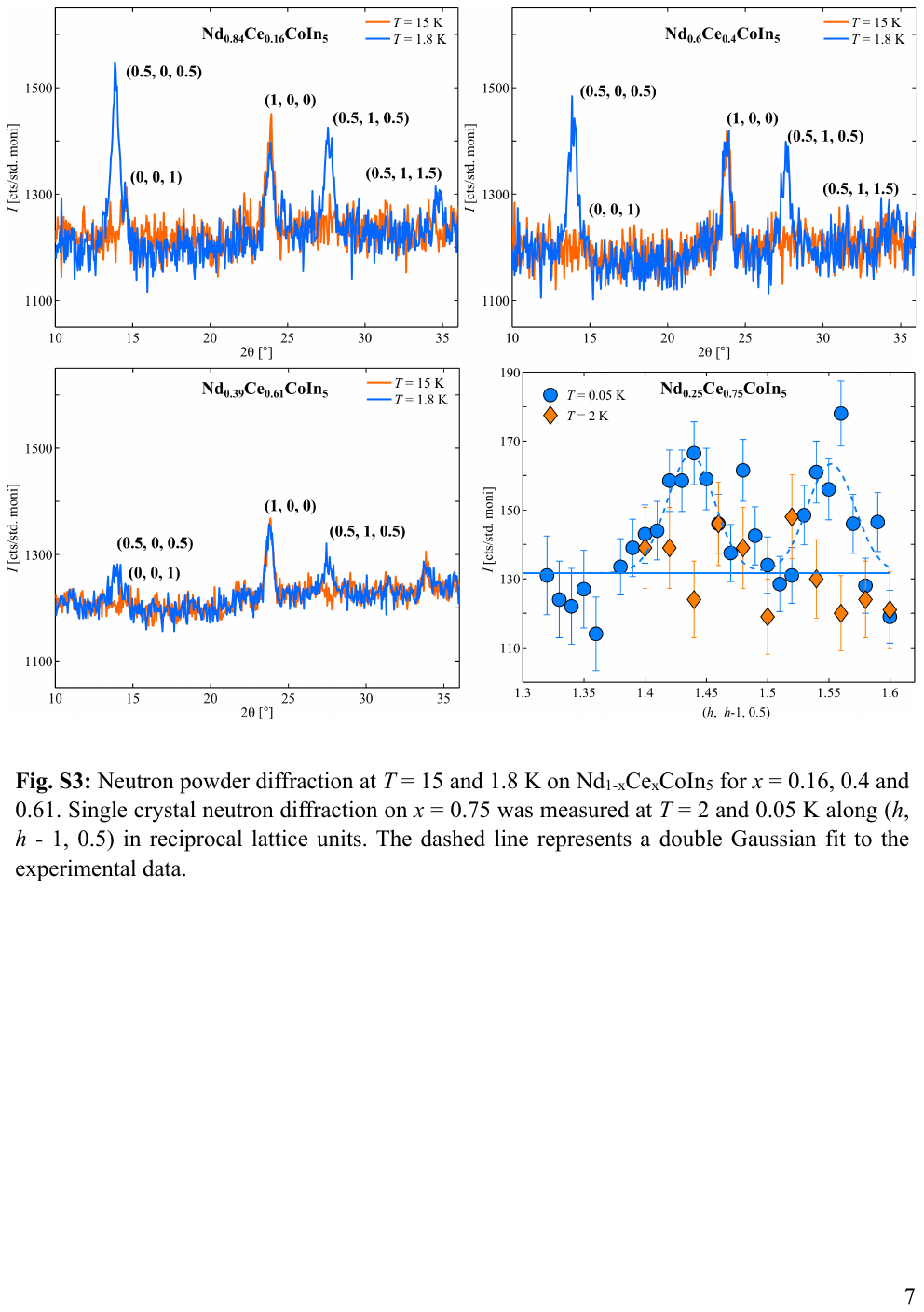}
\end{figure}
\end{document}